\documentclass[]{elsart}

\usepackage{natbib, graphicx}

\usepackage{amssymb}
\journal{Physics Letters A}
\begin{document}

\begin{frontmatter}




\title{ Granular Matter: a wonderful world of clusters in far-from-equilibrium
systems}

\author[liege]{M. Ausloos},
\ead{MarcelAusloos@ulg.ac.be}
\author[liege,bruxelles]{R. Lambiotte},
\author[liege,pol1]{K. Trojan},
\author[liege,pol1]{Z. Koza},
\author[liege,pol2]{M. Pekala}

\address[liege]{SUPRATECS, B5, Sart Tilman Campus, B-4000 Li\`ege, Euroland}
\address[bruxelles]{Universit\'e Libre de Bruxelles, B-1000 Bruxelles, Euroland}
\address[pol1]{Institute of
Theoretical Physics, University of Wroc\l{}aw, pl. Maxa Borna 9, \\PL-50-204
Wroc\l{}aw, Poland}
\address[pol2]{Department of Chemistry, University of Warsaw, \protect\\
Al. \.Zwirki i Wigury 101, PL-02-089 Warsaw, Poland}

\begin{abstract}
In this paper, we recall various features of non equilibrium granular systems.
Clusters with specific properties are found depending
on the packing density, going from loose (a granular gas) to
sintered (though brittle) polycrystalline materials.  The phase space available can be quite different.
Unexpected features, with respect to standard or expected ones in classical fluids or solids, are observed, - 
like slow relaxation processes or anomalous electrical and thermoelectrical transport property dependences.
The cases of  various pile structures and the interplay between classical phase
transitions and self-organized criticality  for avalanches are also outlined.
\end{abstract}


\end{frontmatter}

\section{Introduction}

Granular media are made of a large number of macroscopic solid entities. The
size of these particles ranges from that of dust (in powders) to that of rocks
(in planetary rings). Practically let us note that grains often result from fracture processes
rather than from physico-chemical synthesis.  The granular media well-known features encompass
avalanches and agglomerations,  By extension of the statistical physics point of
view, many other systems can be considered as granular systems, for example car
traffic is
thought to be similar to granular flow.

Transport and storage of granular materials may also be problematic, and
have striking consequences, as in the case of the storage of cereals in silos
that may
break due to the huge internal pressure forces generated by the grains. In
the pharmaceutical industry,  the properties of mixing and segregation of these
systems is primordial  to optimize the contact area between the
chemical components, and consequently the efficiency of the chemical reaction;
moreover, in the  food industry, the properties of
homogenization go on par with the standardization required by mass production.
For these reasons, in the engineering community, there has been a long-standing
interest in describing and
predicting both static and dynamical properties of granular materials.

It is thought that the properties of granular materials mainly originate from
their macroscopic dimensions and surface interactions between the
 grains, which at the relevant scale of granular physics are considered
 as solid objects whose internal degrees of freedom can be neglected. This has
important consequences: the ordinary
 temperature plays no role on the grain motion, and the interaction between
grains is usually dissipative. Yet,
whether the granules in such materials interact through attractive or repulsive
forces is not a trivial question. This is somewhat  unusual physics, and should
have attracted physicist attention.
 
 All this points to a very
rich phenomenology   
which in many cases qualitatively differs from that observed
in the
familiar forms of matter, namely solids, liquids or gases. Over the last decades,
their striking features have been put in parallel with phenomena occurring in
more complex systems, and have been used as a fruitful metaphor to
describe other dissipative dynamical systems. For instance, sand piles have been
used by De Gennes \cite{85} as a macroscopic picture for the motion of flux lines
in superconductors, and by Bak  \cite{8} as a paradigm of the Self-Organized
Criticality (SOC) concept. Other fields of science have borrowed ideas from
granular matter physics, such as crystal agglomeration and sintering, traffic
(jams),
film growth, galaxy clustering, cloud formation (aerosols), company merging
(city planning), fracture. The slow relaxation processes  which take  place in
vibrated sand piles as well as in glasses or flux lattices \cite{149,164} are
other examples. Macroscopic properties, i.e. through mechanical, thermal,
electric or magnetic measurements, can be used to test the internal
structure of piles or compacted matter. Their interpretation remains a
fundamental challenge. Nevertheless it can be expected that macroscopic
truly natural systems will  be described and explained from
microscopic models. This attempt is hereby made under the form of a review of
work by various of us.

In Sect. 2, a few results on Granular Gases will be presented,
though for a more complete presentation, it is best to read the paper by Lipowski
\cite{Lipowskiladek} in these proceedings, and the thesis by Lambiotte
\cite{Lambiottethesis}. The Maxwell Demon experiment will be recalled and it will
be shown that granular gases present complex features such as clustering,
metastability, segregation... The fact that the velocities of the grains do not
distribute themselves like a Maxwellian will also be shortly presented.
Non-equipartition of energy in mixtures and non ergodicity will not be discussed
for lack of space. The Ballistic Magnetic Deposition (BMD) model \cite{BMD1} of
granular piles, somewhat related to the Tetris model \cite{Tetris} will be
discussed in Sect.3. The  role of contact energy, bead anisotropy, and external
field influence will be mentioned in order to warn about problems which might
arise for the stress net estimation and the formation of clusters. The
Bak-Tang-Wiesenfeld (BTW) Sand Pile model \cite{BTW}  will find its place in
Sect. 4, where it will be coupled to the Blume-Emery-Grifiths  (BEG) model
\cite{BEG} in order to test the robustness of SOC. The observation of
log-periodic avalanches in piles built on a fractal basis will be recalled to
have led to a model for the analogous behavior of log-periodic oscillations in
financial indices before an endogenous crash. An anomalous behavior of the
electrical resistivity and the thermoelectric power (TEP) will be recalled in
Sect. 5 for densely packed conducting grains. Conclusions can be found in Sect. 6

\section{Granular Gases} 

\begin{figure}
\begin{center}
\includegraphics[height=4.0in]{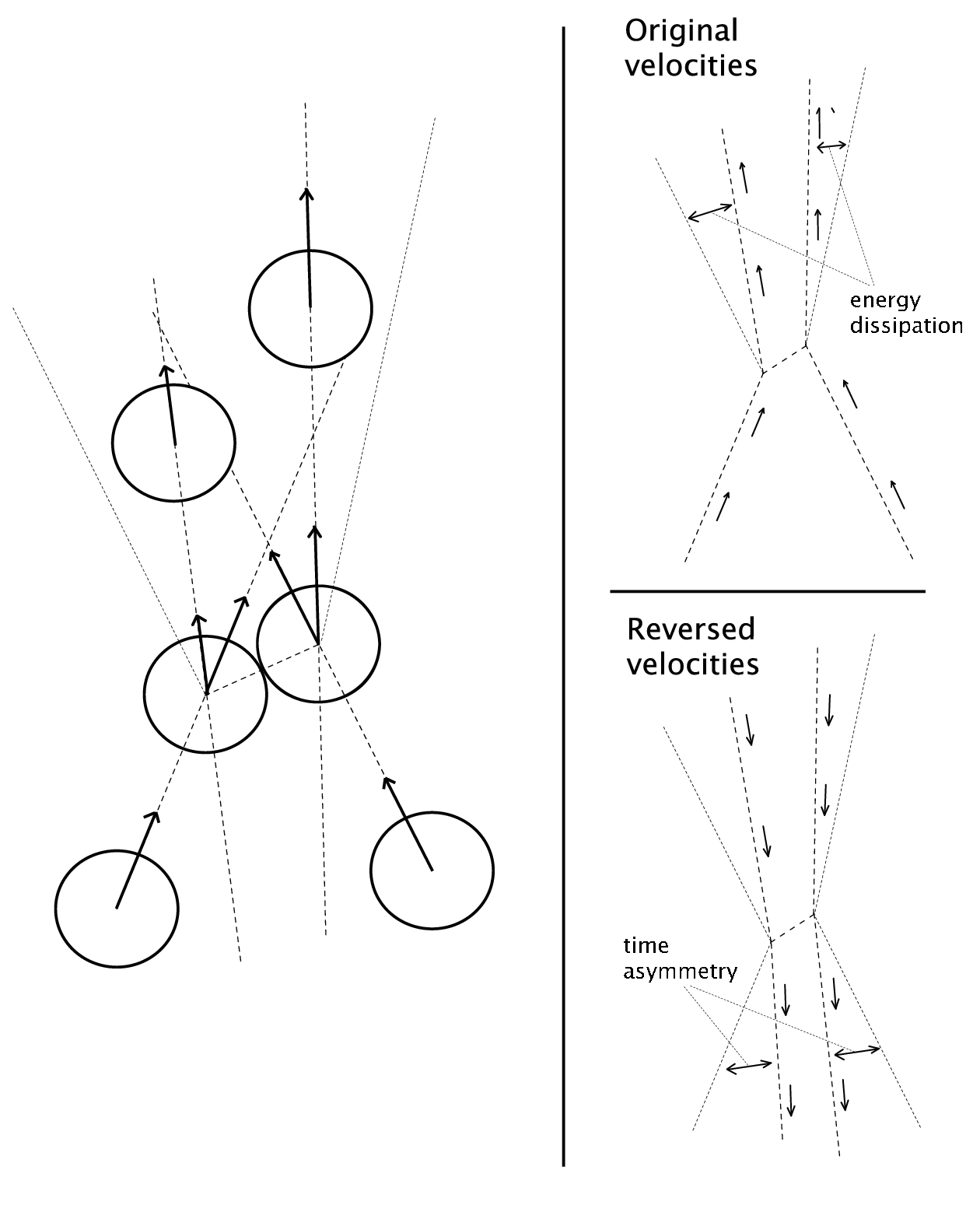} 
\end{center}
\caption{\label{irreversibleInelastic}  Collision between inelastic hard spheres.
Inelasticity  makes the post-collisional velocities more parallel  as compared to
an elastic collision. This is a direct effect of inelasticity which lowers the
quantity $({\mbox{\boldmath$\epsilon$}}.{\bf v}_{12})$ during one collision.
Figures on the right illustrate the irreversibility of the microscopic dynamics:
  the inverted trajectories are no longer the same as the direct
ones}
\end{figure}

In this section we mainly focus on very dilute systems for which the classical
kinetic theory can serve as a basis of investigations. This regime consists in a dilute
assembly of  grains interacting through inelastic scattering processes,  usually
called an inelastic gas. Such  a dilute gas of inelastic hard spheres has been
intensively studied over the last decade in order to highlight the influence of
inelasticity on the properties of a fluid. In this dilute limit,  the inelastic
Boltzmann equation is usually considered in order to study the grains evolution
\cite{ernst2}. As compared to the classical Boltzmann equation, the main
corrections are due to the inelasticity of collisions,  which leads to a
time-irreversibility of collisions and to energy dissipation (Fig.
\ref{irreversibleInelastic}). These effects are taken into account  through the
inelasticity
parameter $\alpha$, which is defined by the relation:
\begin{equation}
\label{relationG}
({\mbox{\boldmath$\epsilon$}}.{\bf v}_{ij})^{*} = -  \alpha  ~
({\mbox{\boldmath$\epsilon$}}.{\bf v}_{ij})
\end{equation} 
where ${\bf v}_{ij} \equiv {\bf v}_{i} - {\bf v}_{j}$, ${\bf r}_{ij} \equiv {\bf
r}_{i} - {\bf r}_{j}$ and
${\mbox{\boldmath$\epsilon$}}$ is a unit vector along the axis joining the
centers of the two colliding
hard spheres $i$ and $j$, and ${\mbox{\boldmath$\epsilon$}} \equiv \frac{{\bf
r}_{ij}}{|{\bf r}_{ij}|}$.  The  velocities with the * index are the  velocities
after the collision and the unleashed symbols correspond to the pre-collisional
velocities. By definition, $\alpha$ belongs to $]0,1]$, and the elastic limit is
recovered for $\alpha=1$.

In \cite{Lambiottethesis}, the dependence of $\alpha$ on the relative velocity of
the particles has been
neglected. This is an approximation since real-world grains have a velocity
dependent on $\alpha$ \cite{ramirez3, brilliantov2}. One should stress that this
simplification has a dramatic but unphysical consequence in MD simulations,  the
so-called collapse phenomenon \cite{mcna2}. This phenomenon, which has been
studied in detail during the last decade, is an extreme expression of the
clustering instability specific to inelastic fluids and consists in the
occurrence of an infinite number of collisions in a finite time.

\begin{figure}
\begin{center}
\includegraphics[width=2.2in]{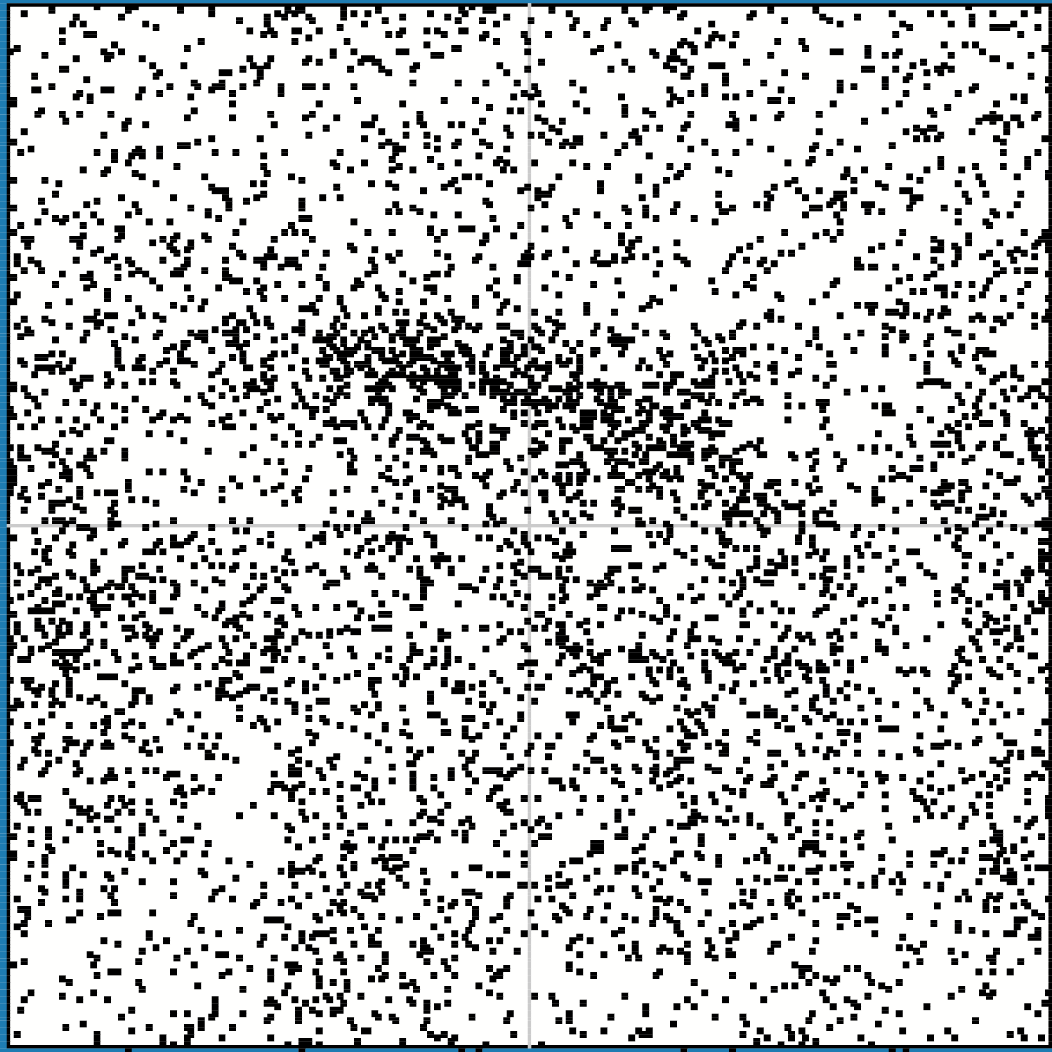}
\hspace{0.1in}
\includegraphics[width=2.2in]{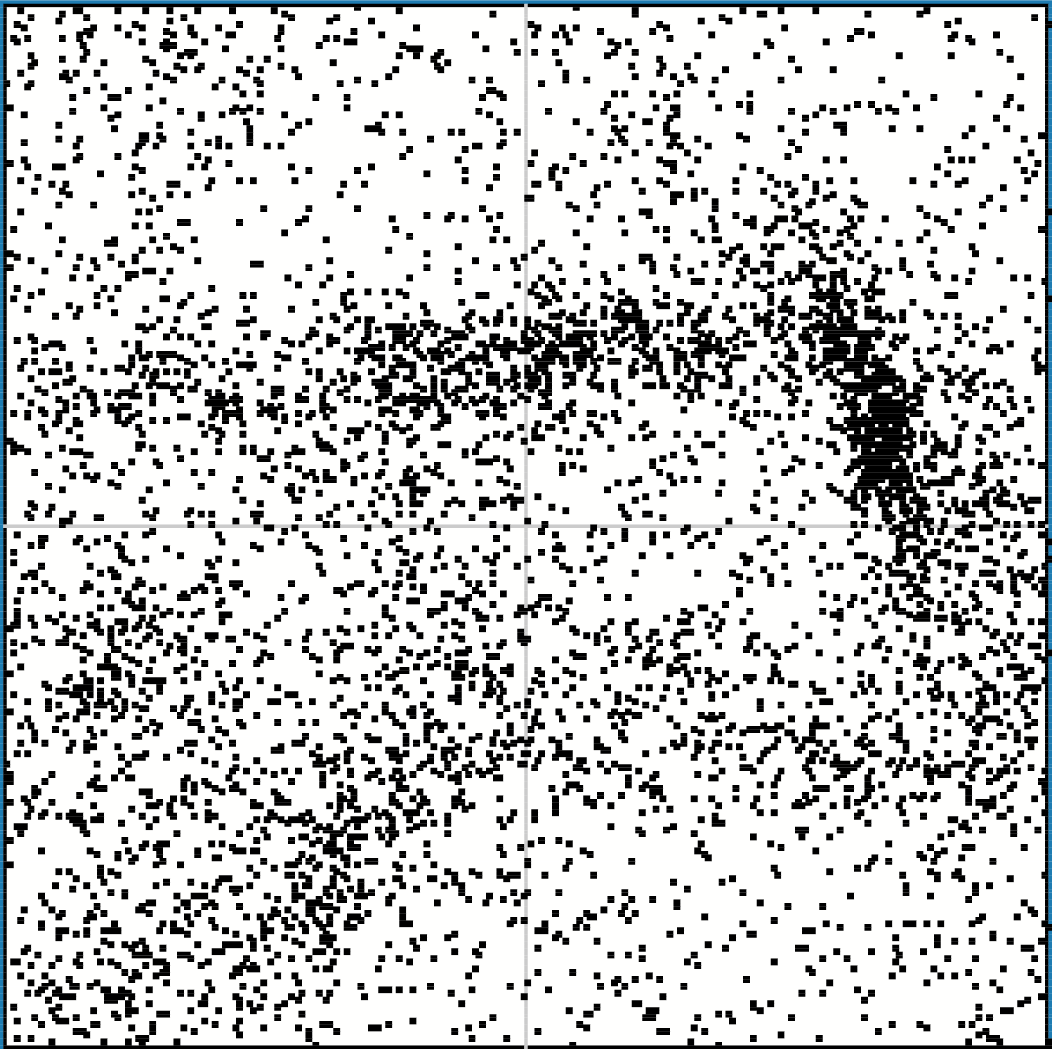} 

\vspace{0.1in}
\includegraphics[width=2.2in]{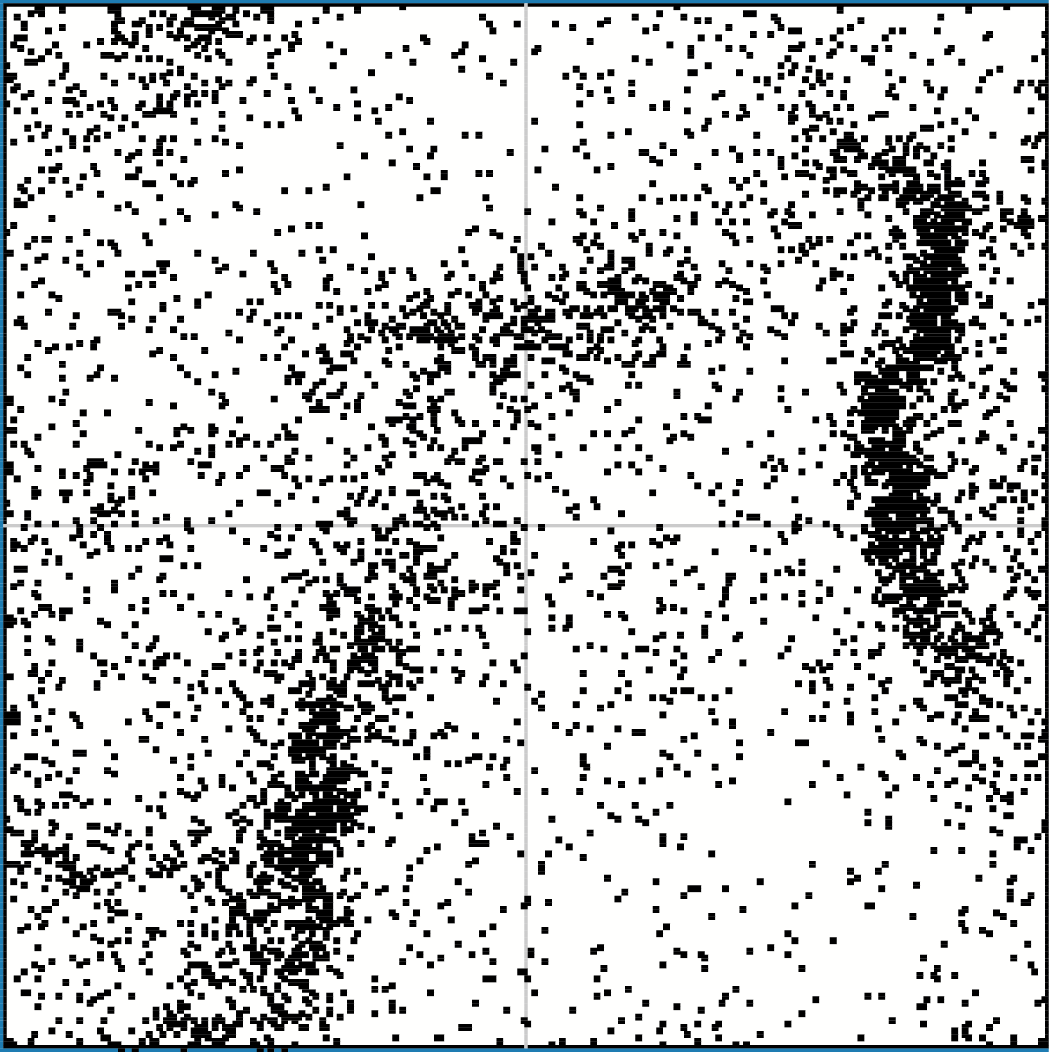}
\hspace{0.1in}
\includegraphics[width=2.2in]{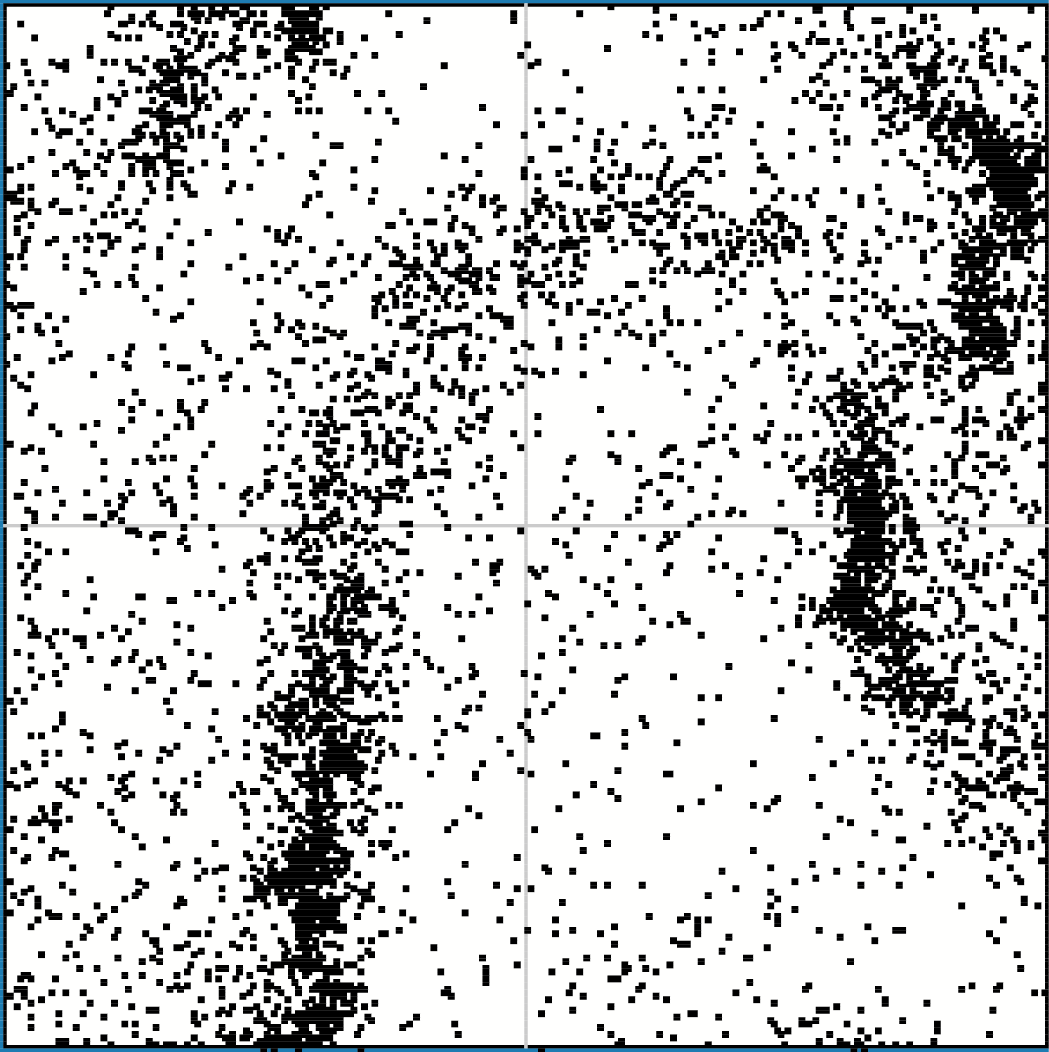}
\end{center}
\caption{\label{clustering}   Aggregation tendency and emergence of clustering in
a freely cooling granular gas. The system is composed of 5000 inelastic discs and
$\alpha=0.9$ in Eq.(1)}
\end{figure}

The dissipativity of collisions contributes to an anomalous exploration of the
phase
space. Indeed, the transfer of the kinetic energy to the internal degrees of
freedom
of the grains makes
the total kinetic energy of the system to decrease and leads to the
irreversibility of the
grain dynamics in  time. This irreversibility favours some
regions in phase space, and generates some complex phenomena, like clustering of
trajectories and heterogeneous density regions (Fig. \ref{clustering}). Moreover,
it also implies that if
no external energy is applied to the system, the system tends toward  a
perfectly resting state,  in which the total kinetic energy vanishes.
Consequently, the exploration of the phase space is slowed down. Before
the system   reaches a complete rest,  it passes through different asymptotic
states,  themselves depending on the initial conditions, because of the lack of
ergodicity which develops in the system. Let us stress that this expression of
metastability does not rest on equilibrium-like mechanisms. The study of
inelastic gases has also led to identifying  the main inelastic effects which
alter the macroscopic dynamics, namely the anomalous coupling between the local
energy and the local density, and the new time scale associated to the
dissipative cooling.

One of the macroscopic consequences of inelasticity is  the emergence of
anomalous transport
processes in inelastic gases. This can be found by applying the Chapman-Enskog
procedure
to the inelastic Boltzmann equation \cite{dufty4}, in order to derive
hydrodynamic equations
for inelastic dilute gases.
 In doing so, a generalized Fourier law for the heat flux of the granular fluid
can be derived:
\begin{equation}
\label{genFourier}
{\bf q}= - \mu \partial_{\bf r} T - \kappa \partial_{\bf r} n.
\end{equation}
Moreover, the mechanisms which make $\kappa \neq 0$ can be identified.  On the
one hand, it is a manifestation of the coupling between density and temperature
which occurs in granular media. Namely, the local temperatures decrease with
different cooling rates depending on the local density. This
non-local behaviour discriminates the temperature dependence of neighboring
hydrodynamic cells and couples the density field to the heat transport process,
leading to an additional term $\sim \partial_{\bf r} n$, and to a positive
definite contribution to $\kappa$. On the other hand, additional contributions to
the heat flux come from the shape of the reference velocity distribution. This
zeroth-order state in the Chapman-Enskog procedure is called the Homogeneous
Cooling State, and consists in a state which is and remains homogeneous, and
whose time dependence occurs only through the granular temperature. It usually
plays the role of the equilibrium state in the statistical description of
granular gases.
This contribution 
can be either negative or positive, depending on the variation to the Maxwellian,
which is usually measured through  
the kurtosis of the velocity distribution.

A striking consequence of this generalized Fourier law is 
the phenomenon of temperature inversion in vibrated granular systems. This
phenomenon occurs in systems subject to gravitation and to which energy is
supplied  by a vibrating bottom wall. 
Because of the energy injection, the inelastic fluid  attains an asymptotic
stationary distribution which is different from the total rest state, and whose
temperature is non-vanishing.
Several authors have shown that the temperature profile is not uniform in the
system \cite{soto1,brey8, meerson1}. Moreover, the granular temperature has been
shown to exhibit a minimum at some distance from the boundary. This fact, which
implies that the granular temperature can increase with increasing height, has been
predicted using Eq. (2), and verified by experiments and numerical
simulations \cite{Lambiottethesis}. A question
remains on whether the Fick and Ohm equations also have to be appropriately
modified  in their own context. Another question pertains to Onsager relations
validity, and whether they should be generalized.

The non-Maxwellian features of velocity distributions arising in inelastic gases
are also an interesting phenomenon.  Indeed, their  tendency toward overpopulated
high energy
tails is non trivial, and the specific shape of the tail  depends on the details
of
the model. 
However, overpopulation seems to be a generic feature of inelastic gases
\cite{ernst2}. This property has been predicted theoretically and numerically,
and observed  experimentally in a large number of situations, see \cite{Lambiottethesis}.
 From a theoretical
point of view,  simplified kinetic models have been introduced  to study the
formation of these fat tails, in pure systems and in mixtures \cite{lam1}. These models, which rest on a mathematical
simplification of the collision operator in the inelastic Boltzmann equation, are
usually called Maxwell Models. Usually, two kinds of asymptotic
velocity distributions are considered: (i) scaling solutions, which
occur when the grains evolve freely, without external forcing and correspond to
 a Homogeneous Cooling State, (ii) heated stationary solutions, which
are obtained by injecting energy into the system to counterbalance the
energy loss. The energy is usually introduced by stochastic forces, usually of
Langevin type. This mimics properties of the vibro-fluidized granular media. 
In \cite{lam2,lam3}, another class of asymptotic solutions has been studied,
namely stationary solutions of the unforced case. By using similarities between
the Maxwell model and a random walk in velocity space, it is shown theoretically
and numerically that almost stationary solutions exist. This means that the core
of these distributions is stationary for an arbitrary  time extent, while the
total energy in the system decreases exponentially fast.
It can be shown that these solutions correspond to truncated L\'evy
distributions.

Finally, let us recall the granular Demon experiment,  conceived to visualize the
energy-density coupling characteristic of granular matter. It consists of a box
divided into two equal compartments by a vertical wall starting from the bottom
of the box, in which a hole allows the passage of the grains from one compartment
to the other. The box is filled with inelastic identical particles submitted to
gravitation. Energy is supplied by a vibrating bottom wall. This simple   system
can be shown to exhibit an order-disorder transition. Indeed, for a high energy
input, the system presents a homogeneous steady state, while when the energy
input is decreased, a phase transition occurs and an asymmetric steady state
prevails. The effusive model proposed by Eggers \cite{Eggers} explains this
transition. In \cite{lam4}, this system has been studied by using a
stochastic urn model of Lipowski, see also \cite{Lipowskiladek,lipo3}. Several
generalizations of the original Maxwell Demon experiment have been considered,
and have led to a rich phenomenology: (i) systems composed by an arbitrary number
of compartments, leading to metastable states. (ii) systems where energy is input
asymmetrically in the urns, showing hysteresis and strong similarities to the
behavior of a ferromagnetic system in a external magnetic field. (iii) the
original experiment applied to mixtures, leading to horizontal segregation and
the emergence of non-stationary oscillating behaviors, called the {\em granular
clock}.

\section{Ballistic Magnetic Deposition Model}

At higher density, granular media lock into piles. 
In constructing and describing granular piles, it is crucial to remember that
basic entities are not made of symmetrical units, like hard spheres; even if they
can
be thought to be made of hard cores \cite{MBE}, like rice grains   or sand grains
\cite{sandpileNagel,sandpileonfractal}, they can be hardly assumed to be
spherical
or disk shaped and perfectly sliding on each other. Specific angles of repose
\cite{reposeangle2}, jams and arches \cite{arches3} are thought to originate from
these contact forces.

We have introduced a so called magnetically-controlled ballistic rain-like
deposition (MBD) model \cite{BMD1,BMD2,BMD3} of granular piles and numerically
investigated its static properties in 2D. The grains are assumed to be elongated
disks characterized by a two-state scalar degree of freedom, called the
''nip''. The direction of a nip can represent e.g. anisotropy of grain, or the
position of a grain with respect to neighboring entities. The nip-nip interaction
is described through the well known Ising-like Hamiltonian.  An external field of
arbitrary origin can be imagined to forcefully allow the grain to rotate during
its deposition. This effect is   introduced through a parameter $q$.

The simulation algorithm creates a pile under a fixed probability $q$ for grain
rotation, or nip change of sign,  during the ballistic grain fall; the
probability to choose the ''up'' direction (or $+1$ value) is $q$.

The local energy gain  $$ E = -J \sum_{<i,j>} n_i n_j $$ is
calculated at each time step since $n_j $ can take the value $+1$ or $-1$ depending on whether
the grain long direction is either vertical or horizontal.

If the ''energy gain'' $\Delta E$ is negative the grain sticks to the cluster
immediately in its ''nip'' state. In the opposite case ($\Delta E$ $>0$) the
grain sticks to the cluster with a probability $\exp (-\Delta E)$ where
$\Delta E$ is the local gain in the energy. If the grain does not stick  
 it continues to fall down. Of course, if the site just below the
grain
is occupied, the grain immediately stops and sticks to the cluster.

\begin{figure}
\begin{center}
\includegraphics[width=\linewidth]{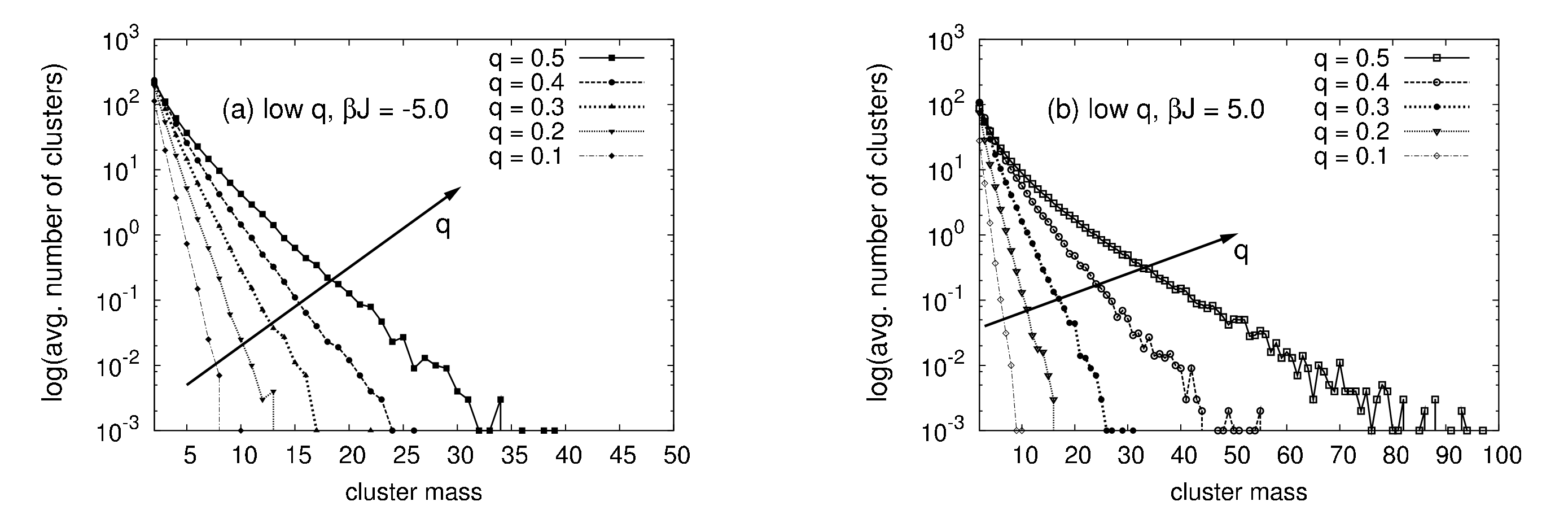}
\end{center}
\caption{\label{fig:ktexp} Semilog plots of the  cluster
size distribution for low $q$
values: (a) $\beta J = -5$, 
(b) $\beta J = 5$. Observe the exponential law-like behavior of the distribution}
\end{figure}

\begin{figure}
\begin{center}
\includegraphics[width=\linewidth]{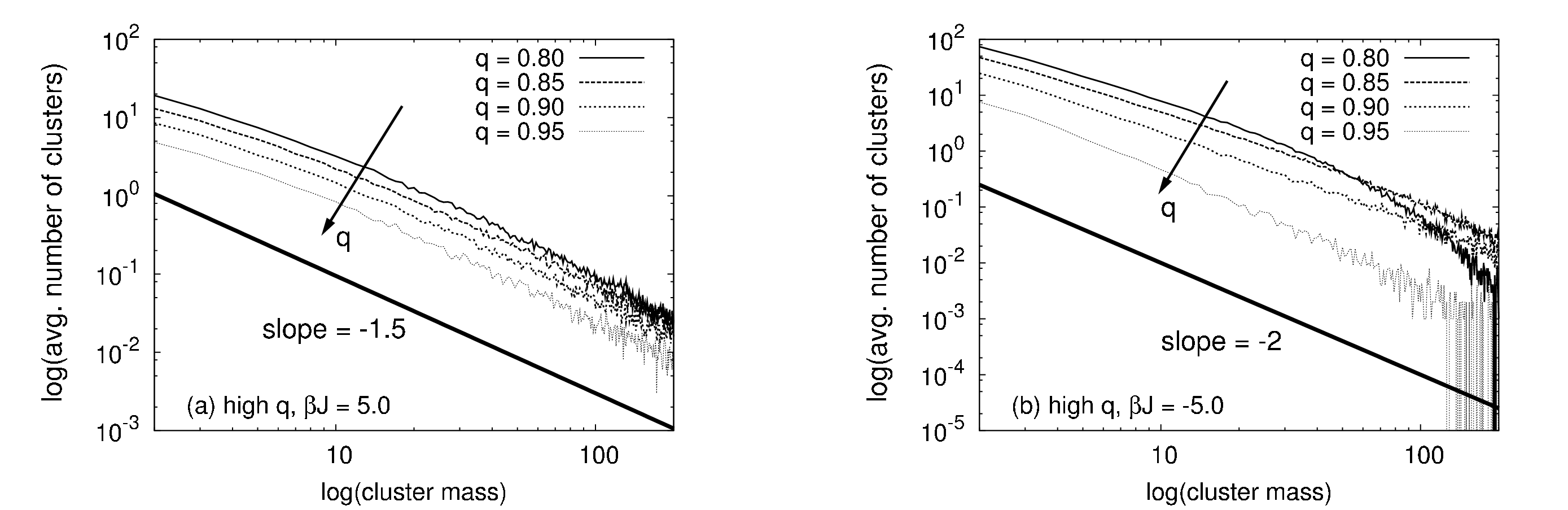}
\end{center}
\caption{\label{fig:ktpow} Log-log plots of the   cluster size distribution for high $q$
values: (a) $\beta J = -5$, 
(b) $\beta J = 5$. Notice the power law-like behavior of the distribution}
\end{figure}

The density and the "$niptization$" (the order parameter for the grain orientation)
have been measured, as well as the fractal-like characteristics of clusters. It is of
interest to report in the present context that clusters of various sizes, with
specific size distributions, occur as a function of $q$ and $J$
(Figs.\ref{fig:ktexp}-\ref{fig:ktpow}).
It has been noticed that the higher the nip-nip interaction strength, the
bigger the difference between the various piles \cite{BMD2,BMD3}.

It seems \cite{BMD3} that one can distinguish pile growth conditions 
and pile structure depending on whether  
  $q<q_c(J)$ or $q>q_c(J)$. Two different cluster-mass regimes have
been identified, through the cluster-mass distribution function. It can be
exponential or follow a power law form depending on whether the nip flip (or grain
rotation) probability $q$ is small or large. It is thought that the regimes are
distinguishable because of a percolation-like transition at finite $q_c$, being
$q_c \simeq 0.85$ and $q_c  \simeq 0.75$, respectively for the $J<0$ and
$J>0$ cases respectively. There is no theory for such features at the moment.

The practical interest is then to calculate the stress net in the pile.  This
is a hyperstatic problem \cite{hyper}. However we have observed \cite{KTMAunpubl} that for
hard spheres the most frequent contact number is $4$.  Therefore, in the case
of  elongated disks the stress net for the BMD model might   be more easily
handled or solved  in two dimensions (2D) than in 3D.
The net has been found to be being unique in 2D. Moreover this contact number value 
 seems
to
remove constraints related to rotation-frustration problems.

Extensions toward binary, polydispersed or more complex objects have not received
much attention. Yet recent studies concerning granular piles suggest the interest
of such further investigations \cite{Herrmann}.

\section{Hybrid Sand Pile Model}

SOC phenomena have recently attracted  much interest.  Their most intriguing
property reside in the `spontaneous' formation of scale-free spatio-temporal
patterns with power-law correlations between various quantities.  On the one hand
these correlations closely resemble those appearing at critical points in
continuous phase transitions; on the other hand, while the critical state at such
 phase transitions is temperature- or external-field-driven, a SOC state is
believed to form spontaneously, without fine-tuning of any external parameters.
The phenomenon basic understanding is still open and the satisfactory definition
of SOC phenomena remains elusive. It was thus of interest to test the robustness
of SOC introducing Hamiltonian like constraints. Whence we  have numerically
investigated a hybrid model, i.e. using the Blume-Emery-Griffiths (BEG)
\cite{BEG} model for the Hamiltonian part, 

 \begin{equation}
    H = -J \sum_{\left<i,j\right>} S_iS_j
        -K \sum_{\left<i,j\right>} S_i^2 S_j^2
	+D \sum_i S_i^2,
    \label{eq:BEG}
 \end{equation}

and the Bak-Tang-Wiesenfeld sand pile
model \cite{BTW} for the SOC part.
The  
Blume-Emery-Griffiths (BEG) model 
is a mere generalization of the above nip model into spin space.
We have considered   a two-dimensional honeycomb lattice of linear size $L$
(Fig.5) . Each lattice node $j$ can be  interpreted
either as
the number of sand grains $h_j\in\{0,1,2\}$ on a site or a spin variable
$s_j\in\{-1,0,1\}$.  Of 6 mappings of heights $h_j$ on spins $s_j$, we choose
the one in which $h_j = 0,1,2$ corresponds to $s_j=0,-1,+1$, respectively.

A node where $h_j>2$ is
called `active', such that
  3 grains are transferred from $j$ to its 3 nearest neighbors, one grain for
each
neighbor.  The process can spontaneously continue, giving rise to an avalanche
of
{\it a priori} unknown size. The whole procedure of adding a grain at a randomly
chosen
node and then completely relaxing, i.e. the resulting avalanche, can be
treated as a Markov process on lattice stable
configurations. It is clear that the set of allowed height (or
spin) configurations is restricted to the set of the recurrent states of this
Markov process, i.e. only the stable lattice configurations that
can be reached arbitrarily many times are considered in this sand pile
'add-and-relax' process.

Sand pile models are known to exhibit strong finite-size effects \cite{Lubeck97}.
This reflects the fact that the system is not uniform---the average density on
the boundary is different than that in the bulk. For
the sand pile part of this hybrid model we have used open boundary conditions (BC).
 However,   for the
Hamiltonian component,
in order to minimize
finite-size effects,  we have adopted periodic BC in our simulations.

Since the phase space of our model is limited to
recurrent states of a Markov process, there is no doubt that
the system is theoretically ergodic. However,  transitions between some states
can
be so infrequent that they will practically never occur in computer
simulations. This problem can be particularly serious for transitions related to
large avalanches.   Any breaking of ergodicity, often related to `glassyness' of a
system,
manifests itself in computer simulations through anomalously slow relaxation.
This has been observed in our simulations.

Nevertheless we have found interesting features.  First,
the high-temperature properties of this hybrid SOC model are very
unusual: while in standard Hamiltonian systems the
magnetisation disappears at high temperatures,
 a nonzero, positive magnetisation $m$ occurs for infinite $T$
 for any values of the control parameters $J$, $K$, and $D$.

It can be argued \cite{ZKMAunpubl} that only two magnetic phases can exist:
one ferromagnetic ($F$) and one antiferromagnetic ($A$) one.
 Note that this
reduction of the number of possible phases is a consequence of the constraints
imposed by the sand pile subsystem, and has nothing to do with the BEG
criticality.

The temperature dependence of the specific heat has been calculated for
different lattice sizes from energy fluctuations,
\begin{equation}
    C = \frac{N}{T^2}\left( \left< u^2\right> - \left< u\right>^2 \right).
    \label{eq:def-C-fluct-E}
\end{equation}

It has been found that a maximum develops in the specific heat
near $T=1.55$, while   $C$ decays as $T^{-2}$ at
 high temperatures.  Moreover very slow relaxation occurs near this temperature,
 as for a second order phase transition, allowing us to call it $T_c$.
 Our results suggest that the effect of slow relaxation dynamics is
amplified by nonlocal constraints of the sand pile model.

One signature of `criticality' of the BTW sand pile model is a power-law decay of
two-point correlation functions $P_{kl}(r)$, of nodes at a distance $r$ apart,
having heights $k$ and $l$ ($0 \le k,l \le 3$). Let $P_k$  be the probability
that a given site has $k$ grains. It was proved that in a two-dimensional
sand pile
\begin{equation}
    P_{kl}(r) = P_kP_l + p_{kl}r^{-4} + \ldots
    \label{eq:Pkl}
\end{equation}
for $k=l=0$ in the bulk \cite{Majumdar91} and for all $0 \le k,l \le 3$ near the
boundary of the system \cite{Ivash94}.

We have examined the case $k=l=0$.
The results for a rather high temperature $T=10$, where the BEG interactions
should not play a significant role, are presented in Fig. 6.
As can be seen, an exponential fit is far better than an algebraic one.
This suggests that the `criticality' of the sand pile model is lost, and the
correlations are dominated by the Hamiltonian subsystem. We believe that
this conclusion remains valid for all finite temperatures.

Conversely we find that the criticality of the BEG model is not affected by the
criticality
of the BTW model. However, even weak interactions of the BEG model destroy
self-organized criticality of the BTW model.

In the above study, the lattice basis is a regular structure.  If the basis
has some fractal structure it has been shown elsewhere that the avalanches
present an extra feature, a log-periodicity, depending on the
fractal dimension and connectivity of the lattice \cite{sandpileonfractal}. Translated into a
complex fractal dimension \cite{Sornette} this sandpile model has then served as an analogy for
describing endogenous financial crashes \cite{Nikkeicrash}.

\section{Conducting Densely Packed Matter}

There is to our knowledge very few papers dealing with the temperature dependence
of electrically conductive packed materials. The electrical properties of such
systems are of interest because there is often a strong modification of the
interface, when an electric current is imposed, - leading to a strong
modification of the oxide layers usually formed at the grain surfaces
\cite{Dorbolo1,Dorbolo2}, even up to welding.

Such systems are usually thought to  behave like a disordered resistor network
characterized by electronic conduction occurring in a strongly localized regime.
Whence transport properties depend whether  the system is above a percolation
transition or not. Carrier hopping between grains dominate the resistance
behavior. Whence such systems display certain features of the
variable-range-hopping (VRH) phenomenon type observed in doped semiconductors. In
particular, the electrical resistivity  of such systems has been observed to obey
a so called  ''fractional temperature dependence'', \begin{equation}
\rho(T)=\rho_0 exp[ (\frac{T_0}{T})^p], \end{equation}  where $T$ is the
temperature, $T_0$  a characteristic temperature, and $p=1/(d+1)$, in terms of
$d$ the dimensionality of the system \cite{Mott1,Mott2,SE1}.

It might be that  by avoiding the oxide layer the electrical conduction mechanism
will also probe the internal structure of the grain, but the more so the
intergrain contact might not conform to the VRH mechanism.

Therefore we have synthesized and compacted a crystalline granular metallic
system,i.e. C$\lowercase{a}$A$\lowercase{l}$$_{2}$. We have used an original
route \cite{Caal2paper} and have obtained tiny crystals. The chemical analysis
has indicated a complex microstructure inherent to the phase diagram intricacies
\cite{Ozturk}. An EDX analysis indicates that the system is made of dendrites
embedded in a matrix. The dendrites are $Al$ rich and made of $CaAl_{2}$ while
the matrix is $Al$ poor and has a composition close to $Ca_{0.6}Al_{0.4}$, - in
fact a mixture of $Ca_{13}Al_{14}$ and $Ca_8Al_{3}$ according to the phase
diagram. These phases have already been studied \cite{HuangCorbett} and have been
found to be metallic with a very similar electrical resistance.

The electrical resistivity and the thermoelectric power of the packed crystal
have been measured under standard conditions. Both properties show three regimes
as a function of temperature (Figs. 7-8). It should be pointed out first that the
electrical resistivity continuously decreases between 15 and 235 K with various
dependences. The  electrical conduction process is found to have a temperature
behavior
$\rho(T)$ $\simeq$ $T^{-3/4}$ at low temperature (Fig.7). This is best
interpreted in
terms of a $thermal$ effect, on a resistive (geometrically disordered) backbone,
not changing with temperature, - in this low $T$ range. The 60-70 K break (or
crossover) indicates the energy range at which the thermal process takes over on
the geometric disorder. At higher temperature,  the smooth decay of $R(T)$ with
increasing $T$ can be attributed to the Fermi-Dirac function behavior increasing
width as $T$ increases and further charge carrier delocalization.

The (positive) thermoelectric power \cite{Caal2paper} shows a bump near 60K,
after an unusual
square root of temperature dependence at low temperature (Fig.8). Above 100 K,  
a
log-log plot indicates that $Q(T)$ behaves like $T^{3/4}$, thus not quite
linearly as it should be  for metallic systems at high  $T$. Recall that  since a
TEP measurement implies no external electrical current. it is unlikely that some
''barrier ageing'' or ''hot spots'' take place in the temperature range of
interest. Within our line of thought, both increase in charge and heat transport
as a function of temperature and the large thermoelectric effect at room
temperature can thus be understood as resulting from a delocalization process on
the intricate barrier network, with   competing characteristic mean free paths,
and a weakening of the contact TEP due to Fermi surface widening with
temperature.

Therefore the three observed regimes are thought to indicate a competition
between geometric and thermal processes in weakly conducting clusters of  densely
packed granular matter. More work is  surely to be done in order to verify the
above
observed behaviors in
other cases and to relate the macroscopic transport properties to the internal
states of the system.

\section{Conclusions}

Granular systems have been considered as a part of applied physics and
technology. Except for some notable attempts by Faraday and Coulomb, they have
not reached the attention of the physics community until recently.

There are several reasons for this new interest. A first reason arises from the
apparent simplicity of granular matter, which surprisingly leads to a very rich
and often counterintuitive behavior. Entropy considerations may be out-weighted
by dynamical effects while the exploration of phase space is unusual rending the
system non-ergodic. The dissipativity of collisions  contributes to this
anomalous exploration of phase space. Indeed, the loss of kinetic energy causes a
decrease of the total  energy of the system, and leads to irreversibility of the
grains dynamics. The simplicity of granular materials originates from the contact
forces acting between their components.

We have first focused on  dilute systems, so that the methods of kinetic theory
may be generalized. In this regime, a dilute assembly of inelastic grains,
usually called an inelastic gas, has been much studied over the last decade in
order to highlight the influence of inelasticity on the properties of a fluid.
Amongst others, it has been used in order to provide a kinetic foundation to
hydrodynamic-like equations for these systems. We have recalled their
characteristic tendency to clustering. The study of inelastic gases has also led
to an identification of the main inelastic effects which alter the macroscopic
dynamics, namely the coupling between the local energy and the local density in
these systems, a  new term in Fourier equation of heat conduction, and a new time
scale associated to the dissipative cooling.

Next we have  outlined a model of gravity deposition for anisotropic 2D gains, in
a more simple way: the gains can either be with their long axis vertical or
horizontal. We have observed that according to the probability of   one  
orientation being more favorable and with a condition of sticking based on an Ising
like Hamiltonian different types of clusters emerge, - distributed in different
ways around a possible percolation like transition.

Next we have observed the interplay of  SOC and Hamiltonian constraints.
Finally we have recalled anomalous  temperature behaviors of dense granular
electrically conducting clusters.

{\bf Acknowledgments}

\vskip 0.6cm

MA and KT thank the organizers of the 41th Karpacz school  conference for their
welcome, stimulating discussions and comments. MA, RL and KT thank also ARC
02-07/2093 for
financial support. MA and MP acknowledge financial support from NATO and CGRI-KBN
bilateral exchange agreements. Paulette CLIPPE and Andrzej PEKALSKI should be
much thanked. 

\newpage 

{\large \bf Figure Captions} \vskip 0.5cm

\begin{figure}
\includegraphics[width=0.9\columnwidth, clip=true, angle=90]{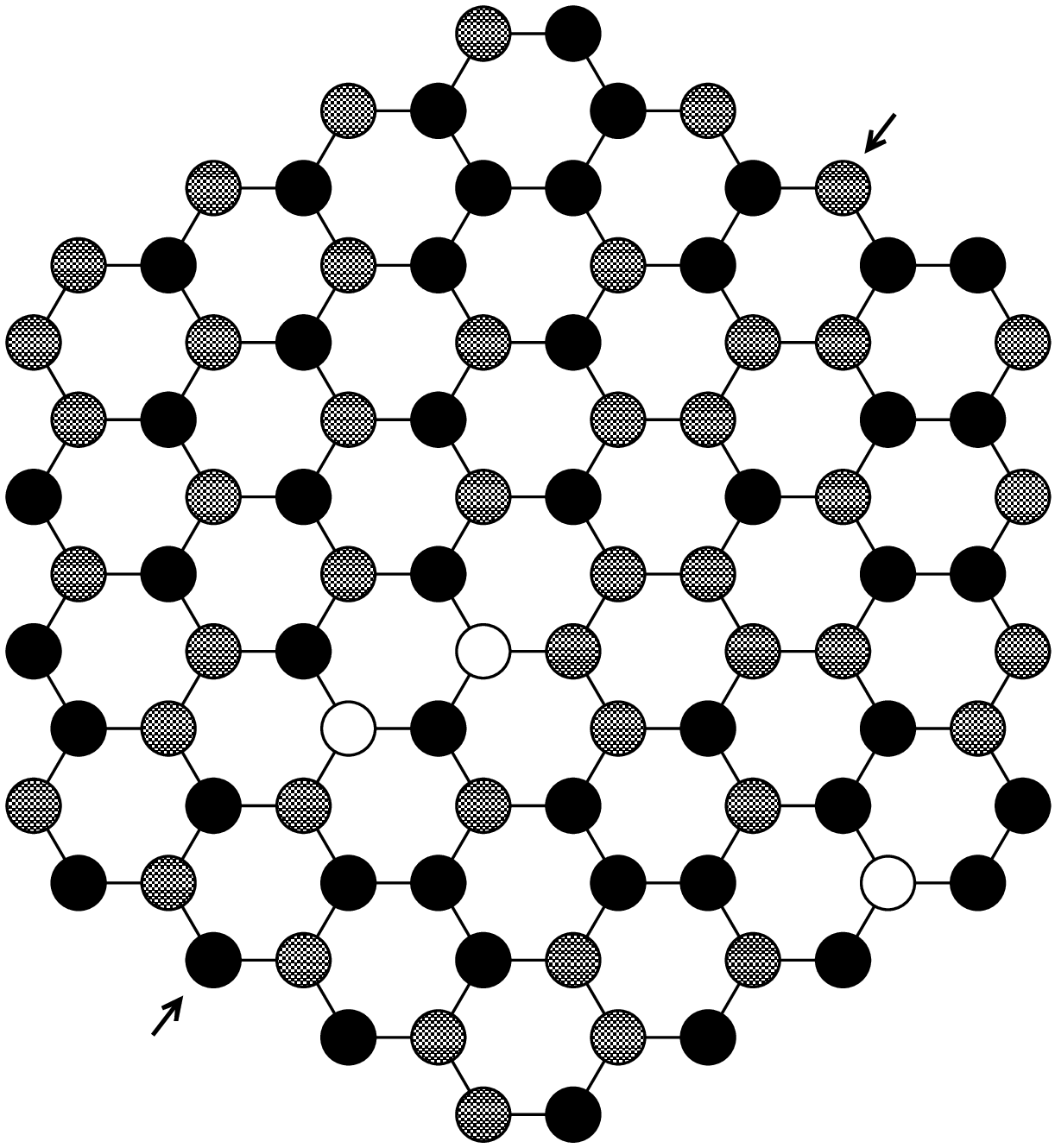}
    \caption{An example of an allowed configuration on a honeycomb lattice 
    of linear size $L=4$. Empty, shadowed, and filled circles represent the nodes with 
    heights $h_j = 0, 1, 2$ (or spins $s_j = 0,-1,+1$), respectively. 
    The arrows show an example of a pair of boundary 
    nodes that interact via Hamiltonian (\protect\ref{eq:BEG}). 
    \label{fig:5}
    }
\end{figure}

\begin{figure}
\includegraphics[width=\columnwidth, clip=true]{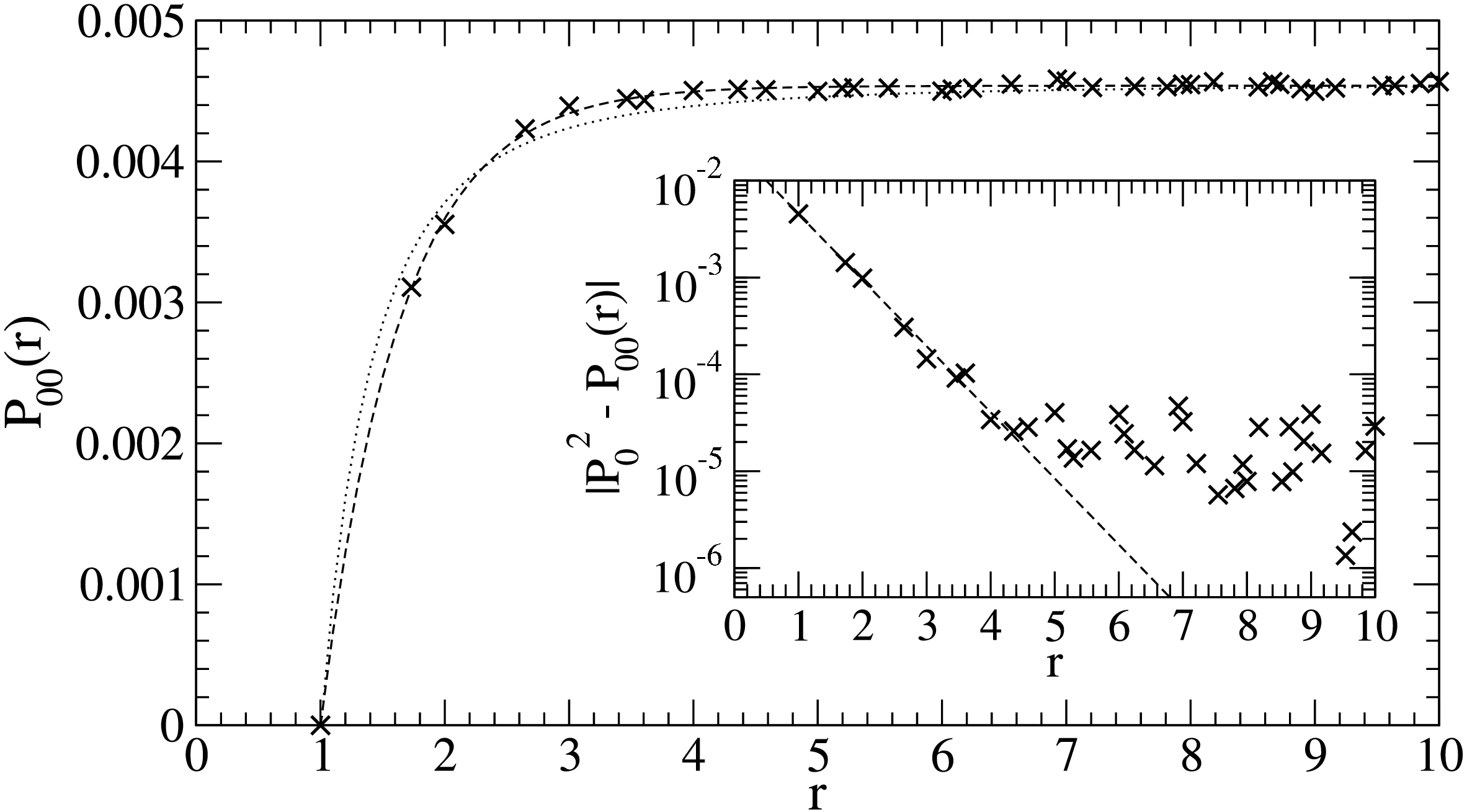}
\caption{Probability $P_{00}(r)$ of finding two empty lattice
   nodes $r$ units apart ($r=1$ corresponds to the distance between nearest
   neighbor nodes). The data were obtained for $L=40$ and $T = 10$. The
   dashed line is an exponential fit (of the form $ a_0 +
   a_1\exp(-a_2r)$), and the dotted line presents a power-law fit ($a_0 +
   a_1r^{-a_2}$). The inset presents the semi-log plot of  $(P_0)^2 - P_{00}(r)$
   and shows that exponential fit is based on 9 points and includes 2 decades on
   the ordinate axis.
    \label{fig:fig_hcorels}
    }
\end{figure}

\begin{figure}
\includegraphics[width=0.9\columnwidth, clip=true, angle=270]{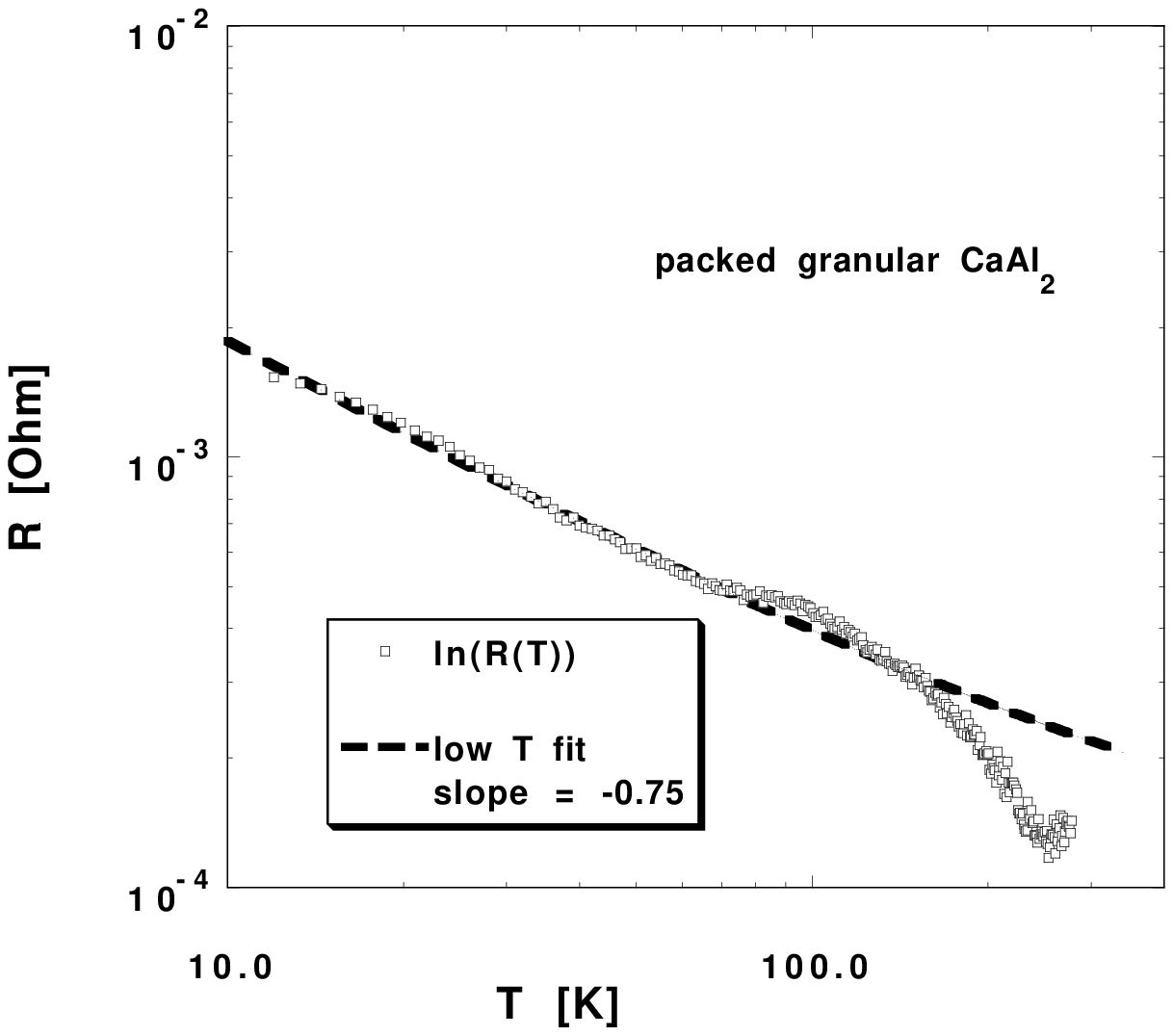}
    \caption{ Electrical resistance $R$ vs. temperature $T$
of a
densely packed  granular CaAl$_{2}$ on a log-log plot, with  a fit
  to a line with slope=-3/4 at low  temperature
    \label{fig:7}
    }
\end{figure}

\begin{figure}
\includegraphics[width=0.9\columnwidth, clip=true, angle=270]{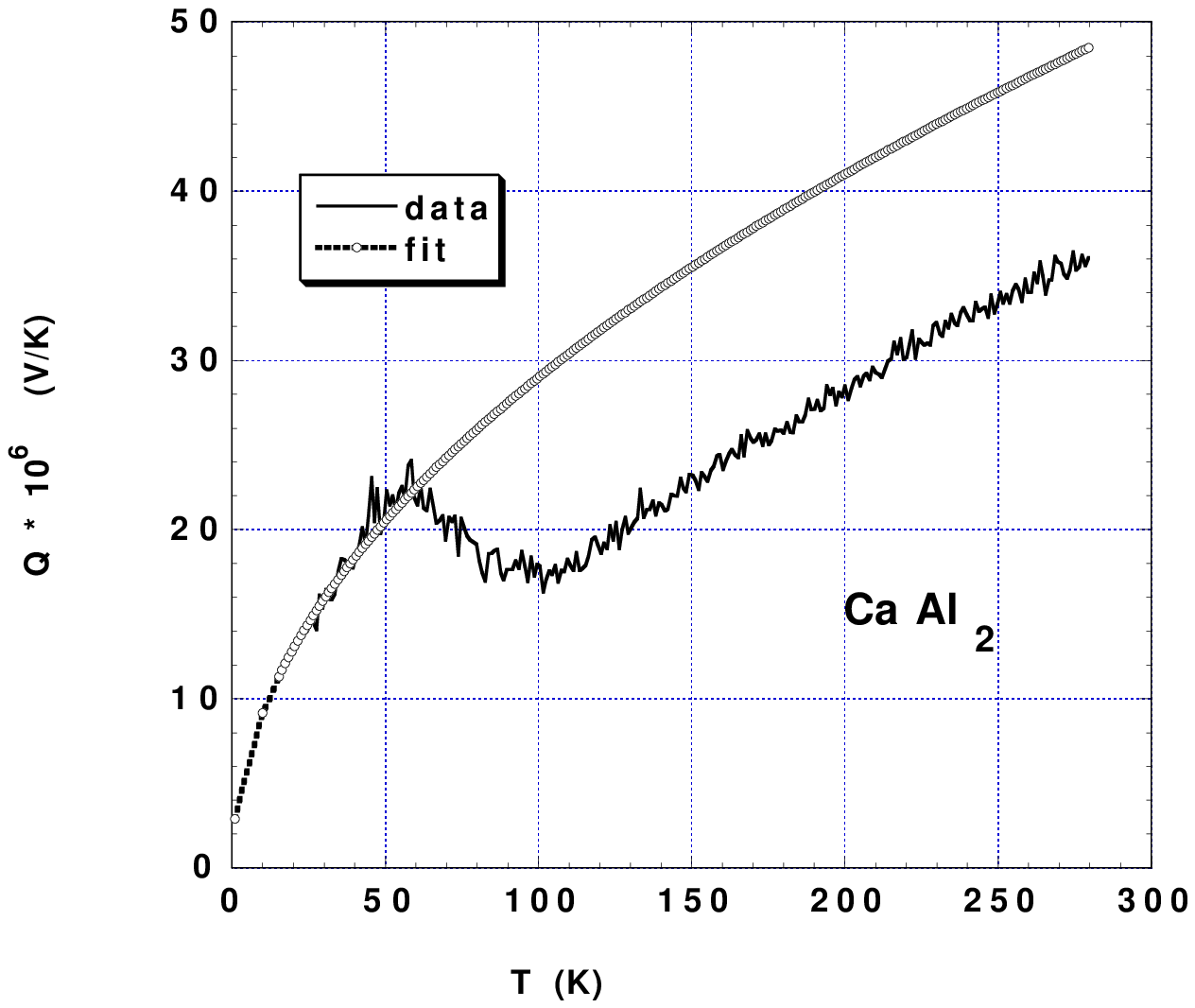}
    \caption{ Thermoelectric power $Q(T) $ vs. temperature $T$ of a densely packed  granular CaAl$_{2}$   with   a square root fit at low  temperature}
    \label{fig:8}
\end{figure}

\end{document}